\documentstyle[12pt,epsf]{article}  
\begin{document}
 
\centerline{\Large Cranked Relativistic Mean Field Description of} 
\vspace{0.5cm}
\centerline{\Large Superdeformed Rotational Bands} 
\vspace{0.5cm}
\centerline{A.V. Afanasjev$^{1,2}$, G.A. Lalazissis$^{1,3}$ and P. Ring$^{1}$} 
\vspace{0.5cm}

\centerline{$^1$ Physik Department der Technischen Universit\"at M\"unchen}
\centerline{D-85747, Garching, Germany}

\centerline{$^2$ Nuclear Research Center, Latvian Academy of Sciences}  
\centerline{LV-2169, Salaspils, Miera str. 31, Latvia}

\centerline{$^3$ Department of Theoretical Physics, Aristotle University of
Thessaloniki,} 
\centerline{GR-54006, Thessaloniki, Greece}
 
\vspace{0.5cm}
\begin{minipage}[h]{14.8cm}
\setlength{\baselineskip}{0.2cm}

{\bf Abstract} {\it The cranked relativistic mean field theory is applied
for a detailed investigation of eight superdeformed rotational 
bands observed in $^{151}$Tb. It is shown that this theory is
able to reproduce reasonably well not only the dynamic 
moments of inertia $J^{(2)}$ of the observed bands but 
also the alignment properties of the single-particle orbitals.} 

\end{minipage}
\vspace{0.5cm}

In the relativistic mean field (RMF) theory the nucleus is described as a
system of point-like nucleons, Dirac spinors, coupled to the
meson and Coulomb fields.
The nucleons interact via the exchange of several mesons, namely a
scalar $\sigma $ -meson, which provides a strong intermediate range
attraction, the isoscalar-vector $\omega $-meson responsible for a very
strong repulsion at short distances and the isovector-vector
$\rho $-meson which
takes care of the symmetry energy. This theory with only seven parameters
fitted to the properties of several spherical nuclei provides an
economic and accurate way to describe many properties of finite nuclei
throughout the periodic table \cite{R.96}.

The RMF theory formulated in the rotating frame - cranked RMF theory \cite
{KR,KR.93} (further CRMF) - has been recently applied for a systematic
investigation of superdeformed (SD) rotational bands observed in the $A\sim
140-150$ mass region \cite{AKR.96}. It was shown that this theory provides
a rather good agreement with the available experimental data on
the dynamic moments of inertia $J^{(2)}$. It reproduces the
trend of the changes of the charge quadrupole
moments $Q_0$. Moreover, the classification of the SD bands in terms of 
the number of filled high-$N$ intruder orbitals, originally suggested within 
the cranked Nilsson (further CN) model \cite{TRA.88}, is supported 
by the CRMF theory.

As the linking transitions from superdeformed states have not been
identified in this mass region, the relative properties of different SD
bands play an important role in our understanding of their structure. One 
way to identify the single-particle orbital by which two SD bands
differ is to compare the difference in their
dynamic moments of inertia $J^{(2)}$ observed in experiment with
the ones obtained in calculations.
However, the deficiency of this approach is that different (especially,
non-intruder) orbitals have rather similar contributions to the 
total $J^{(2)}$. This prevents a unique definition of the underlying 
configuration in terms of non-intruder orbitals due to the 
uncertainty related to the single-particle energies in the SD 
minimum.

An alternative way to analyse the contributions coming from
specific orbitals and thus to 
identify the configuration is the effective alignment approach
suggested by I. Ragnarsson \cite{R.91}. The effective alignment of two bands
is defined as
the difference between their spins at constant rotational frequency
$\Omega _x$: $i^{B,A}_{eff}(\Omega _x)=I_B(\Omega _x)-I_A(\Omega _x)$.
The notation simply reflects the fact that the
spins of the bands under consideration are not experimentally 
determined. In comparison with the method based on $J^{(2)}$, this approach
allows also to study the absolute values of the relative alignment of two SD
bands. This  provides
additional information when theory and experiment are compared. 
Therefore, the application of this approach should be considered
as a necessary test which shows whether theory is able 
to describe the alignment properties of the single-particle
orbitals in the SD minimum.

A systematic investigation of the alignment properties of the 
single-particle orbitals in the SD minimum of the $A\sim 140-150$
mass region has been performed within the CRMF theory 
and the results will be presented in
a forthcoming article \cite{ALR.97}. In the present article, 
the results of the calculations for eight superdeformed
bands observed in $^{151}$Tb \cite{Tb151a,Tb151b} are reported using
the two methods mentioned above. The choice of this nucleus is 
motivated by several reasons. First, the number of observed SD bands is 
the largest among all nuclei in this
mass region. Second, the SD bands observed in this nucleus have been studied
in various non-relativistic theoretical models such as the CN model 
\cite{Tb151a,R.93} and the cranked Hartree-Fock approach with Skyrme forces 
\cite{DD.95,BFH.96,A.97}. Thus, a comparison of our calculations with
the ones obtained with these models can be made.
One should keep in mind that the effective alignment approach has
been used mainly within the CN model so far. Third, the pairing
correlations play a minor role at high rotational frequencies 
($\Omega_x>0.5$ MeV) for nuclei close to $^{152}$Dy. As a result,
the pairing correlations are neglected in the calculations. In
this article calculations have been performed using the non-linear
parameter set NL1 \cite{NL1}. Details of the calculations will 
be given in \cite{ALR.97}.

\input epsf 
\begin{figure}
\vspace*{2.0cm}
\epsfxsize=0pt
\epsfbox{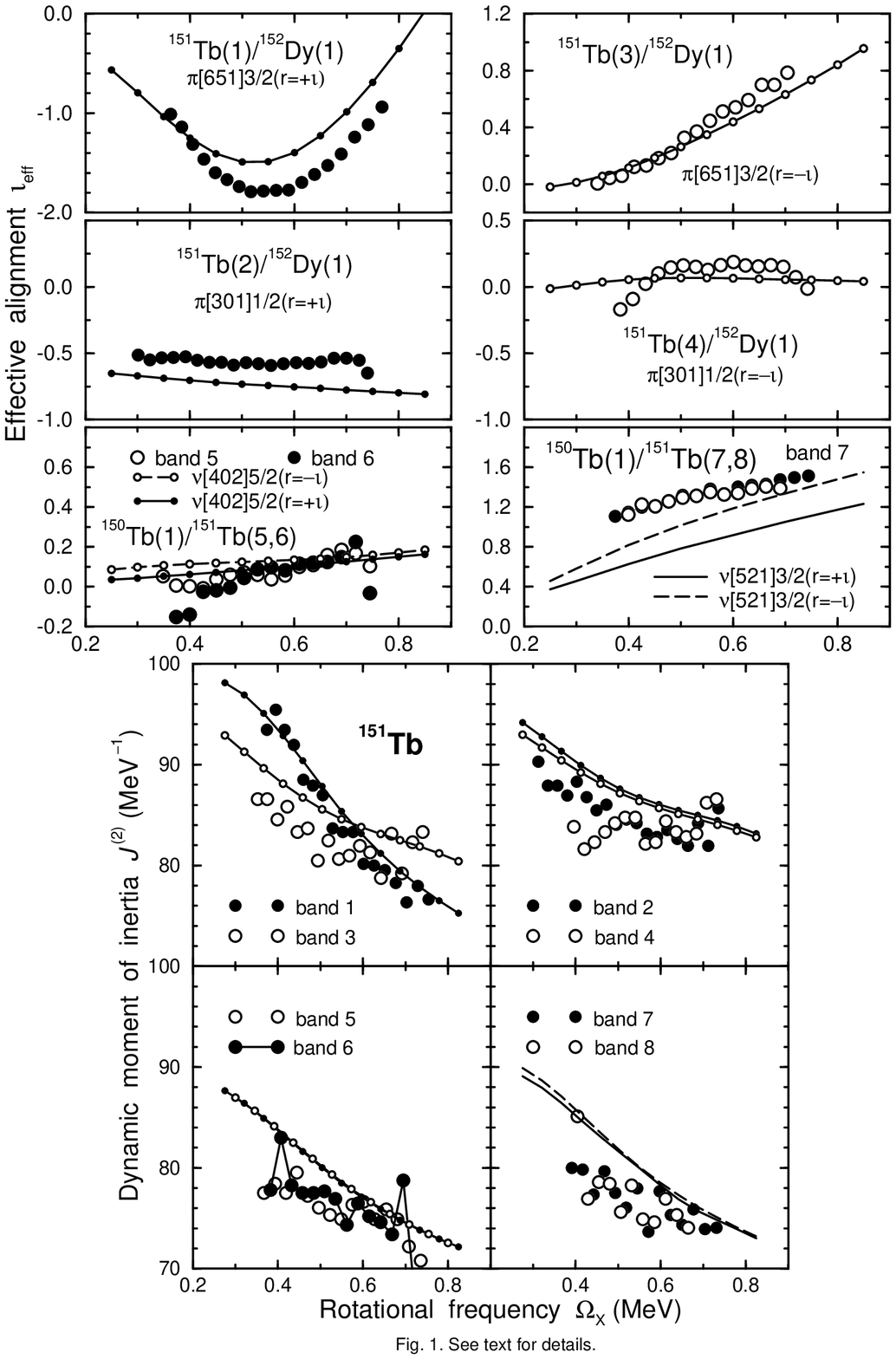}
\end{figure}

In this paper, we restrict ourselves to theoretical configurations
discussed in the literature \cite{R.93,DD.95,BFH.96,A.97}. The results 
of the calculations and the configuration assignment are presented in Fig. 1 
and Table 1. In the upper part of Fig. 1, the effective alignments, $i_{eff}$ 
(in units $\hbar $), extracted from the experimental data (unlinked large symbols) 
and from the assigned calculated configurations (linked small symbols 
of the same type) are shown. The experimental effective alignment between
SD bands A and B is indicated as ``A/B''. The band A in the lighter 
nucleus is taken as a reference so the effective alignment measures 
the effect of the additional particle. 
The experimental data are taken from \cite{Tb151a,Tb151b,Tb150,Dy152}. 
The compared configurations differ by the occupation of the orbital
indicated in the figure. In the lower part of Fig. 1, the experimental 
dynamic moments of  inertia $J^{(2)}$ of observed bands are shown 
along with the calculated ones for assigned configurations. 
The same notation is used as in the upper part.

The interpretation of the bands 1-4 as based on proton holes in the doubly 
magic $^{152}$Dy core discussed in \cite{R.93,DD.95,BFH.96,A.97} 
is consistent with present calculations, see Fig. 1. However, the CRMF
calculations suggest that the spins of the $^{151}$Tb(1) band relative 
to the $^{152}$Dy(1) band should be increased by $2\hbar$ compared with 
the results of the CN calculations \cite{R.93}. 
Fig. 1 shows that the effective alignment of 
the $\pi [651]3/2(r=+i)$ orbital is described much better in the CRMF theory 
than in the CN model, see Fig. 3 in \cite{R.93}.

\begin{table}[ht] 
\caption{Configuration assignment for the bands 1-8 observed in
$^{151}$Tb. Their detailed structure is given relative to the doubly
magic $^{152}$Dy core.
The energies of intraband
$\gamma$-transitions $E_{\gamma}^0$ populating the lowest observed SD 
states are listed in column 3. The 
difference in assigned spins $\Delta I_0$ of the lowest states 
of observed SD bands in $^{151}$Tb and lowest state of yrast SD band 
in $^{152}$Dy is shown in column 4. The difference $\Delta Q_0=
Q_0(^{151}{\rm Tb(N))}-Q_0(^{152}{\rm Dy(1)})$ in charge quadrupole
moments $Q_0$ calculated at $\Omega_x=0.5$ MeV is given in column 5.
}
\begin{center}
\begin{tabular}{ccccc}
\hline\\
Band & Configuration & $E^0_\gamma $(keV) & $\Delta I_0 (\hbar)$ & $\Delta Q_0$ (eb) \\ \hline
   1 &     2                                        &   3    &   4  &  5       \\ \hline
   1 & $\pi [651]3/2(r=+i)^{-1}$                             & 726.5  & +6.5 & $-0.99$   \\ 
   2 & $\pi [301]1/2(r=+i)^{-1}$                             & 602.1  & +0.5 &  +0.20    \\ 
   3 & $\pi [651]3/2(r=-i)^{-1}$                             & 681.5  & +3.5 &  $-0.90$ \\ 
   4 & $\pi [301]1/2(r=-i)^{-1}$                             & 768.6  & +7.5 &  +0.24    \\ 
   5 & $\pi [651]3/2(r=+i)^{-1} $ & 709.3  & +6.5 & $-1.80$  \\ 
     & $\nu [770]1/2(r=-i)^{-1} \nu [402]5/2(r=-i)^1 $ & &  &   \\ 
   6 & $\pi [651]3/2(r=+i)^{-1}  $ & 739    & +7.5 & $-1.80$  \\ 
     & $\nu [770]1/2(r=-i)^{-1} \nu [402]5/2(r=+i)^1 $ &     &  &   \\ 
   7 & $\pi [651]3/2(r=+i)^{-1} $ & 758    &  +9.5    & $-1.51$  \\
     & $\nu [770]1/2(r=-i)^{-1} \nu [521]3/2(r=+i)^1  $ &   &      &   \\
   8 & $\pi [651]3/2(r=+i)^{-1}$                    & 785   & +10.5     & $-1.51$   \\
     & $\nu [770]1/2(r=-i)^{-1} \nu [521]3/2(r=-i)^1$ &    &      &    \\ \hline
\end{tabular}
\end{center}
\end{table}

\vspace{0.2cm}

The effective alignment of the $^{152}$Dy(1) band relative to the bands 5-8
in $^{151}$Tb is strongly increasing as a function of rotational frequency
with total gain in $i_{eff}$ of $\approx 3\hbar$. Considering that 
in the vicinity of the $N=86$ and $Z=66$ SD shell gaps only the $\nu
[770]1/2(r=-i)$ orbital has such properties, this indicates that the 
$^{151}$Tb(5-8) bands are most likely based on neutron excitations from the $\nu
[770]1/2(r=-i)$ orbital into the orbitals above the $N=86$ SD shell gap. 
This simple consideration shows that the interpretation of band 7 
discussed in \cite{A.97} as based on proton excitation across 
the $Z=66$ SD shell gap is rather unrealistic. 

It was suggested in \cite{Tb151a} that bands 5 and 6 are most likely 
based on neutron excitations from $\nu [770]1/2(r=-i)$ to 
$\nu [402]5/2(r=\pm i)$. 
Similarly to the CN model,
see Fig. 3 in  \cite{Tb151a}, the results of our calculations are in good
agreement with experiment which can be considered as an additional
confirmation of this interpretation. The consistent 
interpretation of bands 7 and 8 is more difficult. 
It was  suggested in \cite{Tb151a} that these bands are connected 
with neutron excitations from $\nu [770]1/2(r=-i)$ to 
$\nu [521]3/2(r=\pm i)$.  Although the calculated values of $J^{(2)}$
for configurations built on such excitations 
are reasonably close to the experimental ones, see Fig. 1, an
important discrepancy between experiment and calculations exists 
for the effective alignments in the $^{150}$Tb(1)/$^{151}$Tb(7,8) 
pairs. A strong argument against such an interpretation comes 
from the large signature splitting between the two signatures of the  
$\nu [521]3/2$ orbital. This splitting is rather similar both in the 
CRMF theory and in the CN model and it is in contradiction 
with the experimental data. One should note that the next orbital
above the $N=86$ SD shell gap, namely, $\nu [514]9/2$ 
is signature degenerated and configurations based on
neutron excitations from $\nu [770]1/2(r=-i)$ to 
$\nu [514]9/2(r=\pm i)$ have nearly constant effective 
alignment (which is essentially close to zero) relative to the 
configuration of the $^{150}$Tb(1) band. 

In conclusion, cranked relativistic mean field theory has been applied 
for an investigation of eight SD bands observed in $^{151}$Tb. The calculated
dynamic moments of inertia and effective alignments are in reasonable 
agreement with experiment for bands 1-6. The consistent interpretation 
of bands 7 and 8 in a pure single-particle picture is more difficult. 
This might be connected with the influence of the residual interactions 
neglected in the present calculations. 

\vspace{0.2cm} 
One of the authors (A.V.A.) acknowledges financial support from the 
{\it Konferenz der Deutschen Akademien der Wissenschaften 
through the Volkswagenstiftung} and valuable discussions with
I. Ragnarsson. This work is also supported partly
by the Bundesministerium f\"ur Bildung und Forschung under
the project 06 TM 743 (6).

\end{document}